\documentclass[twocolumn]{aastex63}
\usepackage{amsmath}
\usepackage{booktabs}

\newcommand\oneD{\mathrm{1D}}
\newcommand\twoD{\mathrm{2D}}
\newcommand\Rp{R_\mathrm{p}}
\newcommand\RponeD{R_{\mathrm{p}, \oneD}}
\newcommand\RptwoD{R_{\mathrm{p}, \twoD}}
\newcommand\Rstar{R_{\mathrm{*}}}
\newcommand\M{\mathrm{M}}   
\newcommand\E{\mathrm{E}}   
\newcommand{\appropto}{\mathrel{\vcenter{
  \offinterlineskip\halign{\hfil$##$\cr
    \propto\cr\noalign{\kern2pt}\sim\cr\noalign{\kern-2pt}}}}}

\submitjournal{ApJL}

\shorttitle{Explaining the Cold Retrieved Temperatures of Exoplanet Terminators}
\shortauthors{MacDonald et al.}

\graphicspath{{./}{figures/}}

\begin{document}

\title{Why is it so Cold in Here?: \\ Explaining the Cold 
Temperatures Retrieved from Transmission Spectra of Exoplanet Atmospheres}

\correspondingauthor{Ryan MacDonald}
\email{rmacdonald@astro.cornell.edu}

\author[0000-0003-4816-3469]{Ryan J. MacDonald}
\author[0000-0002-8515-7204]{Jayesh M. Goyal}
\author[0000-0002-8507-1304]{Nikole K. Lewis}
\affiliation{Department of Astronomy and Carl Sagan Institute, Cornell University, 122 Sciences Drive, Ithaca, NY 14853, USA}

\begin{abstract}

Transmission spectroscopy is a powerful technique widely used to probe exoplanet terminators. Atmospheric retrievals of transmission spectra are enabling comparative studies of exoplanet atmospheres. However, the atmospheric properties inferred by retrieval techniques display a significant anomaly: most retrieved temperatures are far colder than expected. In some cases, retrieved temperatures are $\sim 1000\,$K colder than $T_{\rm eq}$. Here, we provide an explanation for this conundrum. We demonstrate that erroneously cold temperatures result when 1D atmospheric models are applied to spectra of planets with differing morning-evening terminator compositions. Despite providing an acceptable fit, 1D retrieval techniques artificially tune atmospheric parameters away from terminator-averaged properties. Retrieved temperature profiles are hundreds of degrees cooler and have weaker temperature gradients than reality. Retrieved abundances are mostly biased by $> 1\sigma$ and sometimes by $> 3\sigma$, with the most extreme biases for ultra-hot Jupiters. When morning-evening compositional differences manifest for prominent opacity sources, H$_2$O abundances retrieved by 1D models can be biased by over an order of magnitude. Finally, we demonstrate that these biases provide an explanation for the cold retrieved temperatures reported for WASP-17b and WASP-12b. To overcome biases associated with 1D atmospheric models, there is an urgent need to develop multidimensional retrieval techniques.

\end{abstract}


\section{Introduction}
\label{sec:intro}

The atmospheric composition and temperature structure of planetary atmospheres are key to understanding the physical processes shaping these worlds. Transmission spectroscopy has proved one of the most successful methods to characterise exoplanetary atmospheres. Observations from the ground and space have yielded detections of various atoms, molecules, and ions \citep[e.g.][]{Deming2013,Sedaghati2017,Spake2018,Hoeijmakers2018}. Sufficiently high-precision observations contain information on terminator temperature structures \citep{Barstow2013,Rocchetto2016}. With over 40 exoplanets now possessing transmission spectra \citep{Madhusudhan2019}, comparative studies of exoplanetary atmospheres are underway \citep{Sing2016,Barstow2017,Tsiaras2018,Fisher2018,Pinhas2019,Welbanks2019b}. A key goal is to measure abundances for a range of volatile species, offering a crucial link to exoplanetary formation mechanisms \citep[e.g.][]{Oberg2011,Piso2016}.

Atmospheric properties can be derived from exoplanet spectra via atmospheric retrieval techniques. Retrievals couple a parametric atmosphere and radiative transfer model to a Bayesian sampling algorithm \citep{Madhusudhan2018}, yielding statistical constraints on model parameters (abundances, temperature, etc.). Precise atmospheric constraints require high-precision observations over a long spectral baseline, including both optical and near-infrared data \citep[e.g.][]{Wakeford2018,Pinhas2019}. However, \emph{reliable} atmospheric inferences further require the retrieval model itself to encapsulate the true nature of the planet under study. For example, a retrieval excluding a molecule which is actually present may arrive at an erroneous solution, despite obtaining a decent spectral fit, by tuning other atmospheric properties away from their real values. 

\begin{deluxetable*}{lclcccclclccc} \label{tab:litvalues}
\tablenum{1}
\tablecaption{Retrieved Temperatures From Optical+IR Exoplanet Transmission Spectra}
\tablewidth{0pt}
\tablehead{
\multicolumn{1}{l}{Planet} & \multicolumn{1}{c}{$T_{\rm eq}$} & \multicolumn{1}{l}{\hspace{5pt} $T_{\rm ret}$} & \multicolumn{1}{c}{$T_{\rm ret} - T_{\rm eq}$} & \multicolumn{1}{c}{$\frac{T_{\rm ret}}{T_{\rm skin}}$} & \multicolumn{1}{c}{Ref.} & \multicolumn{1}{c}{|} & \multicolumn{1}{l}{Planet} & \multicolumn{1}{c}{$T_{\rm eq}$} & \multicolumn{1}{l}{\hspace{5pt} $T_{\rm ret}$} & \multicolumn{1}{c}{$T_{\rm ret} - T_{\rm eq}$} & \multicolumn{1}{c}{$\frac{T_{\rm ret}}{T_{\rm skin}}$} & \multicolumn{1}{c}{Ref.} \\ 
 & (K) & \hspace{5pt} (K) & (K) & & & & & (K) & \hspace{5pt} (K) & (K) & 
}
\startdata \\[-8pt]
GJ~3470b & 650 & 400$^{+100\, \dagger}_{-100}$ & -250 & 0.73 & $^1$ 
& & WASP-127b & 1420 & 820$^{+91}_{-80}$ & -600 & 0.69 & $^8$ \\
`` ''      & 693 & 500$^{+150\, \ddagger}_{-150}$ & -150 & 0.86 & $^{2}$ 
& &`` '' & 1400 & 950$^{+200\, \ddagger}_{-100}$ & -450 & 0.81 & $^{2}$\\
HAT-P-11b & 831 & 750$^{+100\, \ddagger}_{-250}$ & -81 & 1.07 & $^{2}$ 
& &  HD~209458b & 1450 & 1071$^{+149}_{-161}$ & -379 & 0.88 & $^9$ \\
HAT-P-12b & 960 & 456$^{+70}_{-40}$ & -504 & 0.56 & $^3$ 
& & `` ''& 1450 & 949$^{+252}_{-109}$ & -501 & 0.78 & $^3$ \\
 `` ''  & 960 & 610$^{+180\, \ddagger}_{-100}$ & -350 & 0.76 & $^{2}$ 
& & `` '' & 1450 & 950$^{+250\, \ddagger}_{-100}$ & -500 & 0.78 & $^{2}$ \\  
HAT-P-26b & 990 & 550$^{+150}_{-100}$ & -440 & 0.66 & $^4$ 
& &  WASP-31b & 1580 & 1043$^{+287}_{-172}$ & -537 & 0.79 & $^3$\\
`` '' & 990 & 563$^{+59}_{-55}$ & -427 & 0.68 & $^5$ 
& & `` '' & 1580 & 1050$^{+100\, \ddagger}_{-100}$ & -530 & 0.79 & $^{2}$ \\ 
`` '' & 994 & 510$^{+60\, \ddagger}_{-60}$ & -484 & 0.61 & $^{2}$ 
& & WASP-17b & 1740 & 1147$^{+259}_{-305}$ & -593 & 0.78 & $^3$ \\
WASP-39b & 1116 & 920$^{+70}_{-60}$ & -196 & 0.98 & $^6$ 
& & `` '' & 1740 & 1400$^{+200\, \ddagger}_{-200}$ & -340 & 0.96 & $^{2}$ \\ 
`` '' & 1120 & 775$^{+282}_{-166}$ & -345 & 0.82 & $^3$ 
& & WASP-79b & 1800 & 1140$^{+180}_{-180}$ & -660 & 0.75 & $^{10}$ \\
`` '' & 1120 & 1050$^{+100\, \ddagger}_{-100}$ & -70 & 1.11 & $^{2}$
& & WASP-19b & 2050 & 1386$^{+370}_{-337}$ & -664 & 0.80 & $^3$ \\
HD~189733b & 1200 & 1159$^{+146}_{-157}$ & -41 & 1.15 & $^3$ 
&& `` '' & 2050 & 1750$^{+100\, \ddagger}_{-100}$ & -300 & 1.02 & $^{2}$\\
`` '' & 1200 & 775$^{+75\, \ddagger}_{-75}$ & -425 & 0.77 & $^{2}$ 
& & WASP-12b & 2500 & 1455$^{+415}_{-415}$ & -1045 & 0.69 & $^{11}$ \\ 
WASP-52b & 1300 & 630$^{+130}_{-121}$ & -670 & 0.58 & $^7$ 
& & `` '' & 2510 & 990$^{+169}_{-122}$ & -1520 & 0.47 & $^3$ \\
HAT-P-1b & 1320 & 1114$^{+251}_{-205}$ & -206 & 1.00 & $^3$
& &  `` '' & 2510 & 1050$^{+200\, \ddagger}_{-100}$ & -1460 & 0.50 & $^{2}$ \\
`` '' & 1320 & 1075$^{+175\, \ddagger}_{-175}$ & -245 & 0.97 & $^{2}$ 
& & WASP-121b & \hspace{-10pt} $\geq$2500 & 1554$^{+241}_{-271}$ & -946 & 0.74 & $^{12}$ \\[3pt]
\enddata 
\tablenotetext{\dagger}{Estimated from retrieved temperature profile. \hspace{6.5cm} $\ddagger$ Estimated from $T_0$ posteriors.}
\tablecomments{all retrieved temperatures come from studies satisfying the following criteria: (i) both optical and near-IR data are included; (ii) temperatures and chemical abundances are free parameters; and (iii) radiative transfer is numerically evaluated, rather than using semi-analytic approximations. Where a non-isothermal temperature profile is used, $T$\,(1\,$\mu$bar) is quoted. The skin temperature is given by $T_{\rm skin} = 2^{-1/4} \, T_{\rm{eq}}$. The temperature differences and ratios use the median retrieved values.}
\tablerefs{\citet{Benneke2019}$^1$, \citet{Welbanks2019b}$^{2}$, \citet{Pinhas2019}$^3$, \citet{Wakeford2017,Wakeford2018}$^{4,6}$, \citet{MacDonald2019,MacDonald2017}$^{5,9}$, \citet{Bruno2019}$^7$, \citet{Spake2019}$^8$,  \citet{Sotzen2020}$^{10}$, \citet{Kreidberg2015}$^{11}$, \citet{Evans2018}$^{12}$.}
\vspace{-10pt}
\end{deluxetable*}

Here, we draw attention to an anomaly which has emerged from retrieval studies of transmission spectra: \emph{almost all retrieved temperatures are notably cooler than planetary equilibrium temperatures}. In Table~\ref{tab:litvalues}, we summarise inferred temperatures of exoplanet terminators from the literature. The retrieved temperatures of most hot Jupiters are seen to reside $\sim 200 - 600$\,K cooler than $T_{\rm eq}$, whilst for ultra-hot Jupiters this increases to $\gtrsim 1000$\,K. Stratospheric temperatures are not expected to be much cooler than the skin temperature, $T_{\rm{skin}} = 2^{-1/4} \, T_{\rm{eq}}$ \citep[e.g.][]{Barstow2013}, however, 11 of the 16 planets in Table~\ref{tab:litvalues} have median retrieved temperatures colder than their skin temperature. We focus on studies including both optical and infrared data, free chemical abundances, and numerical radiative transfer. This ensures any trends are unbiased by limited wavelength coverage \citep{Wakeford2018,Pinhas2019}, equilibrium chemistry assumptions, or semi-analytic approximate methods \citep[see][]{Welbanks2019a}. We note that infrared only retrievals have also reported anomalously cold temperatures \citep[e.g.][]{Kreidberg2014,Tsiaras2018}; here we focus on the subset also including optical data to mitigate potential biases arising from the absence of a spectral continuum \citep[cf.][]{Heng2017}.

\begin{figure*}[ht!]
    \centering
    \includegraphics[width=\textwidth]{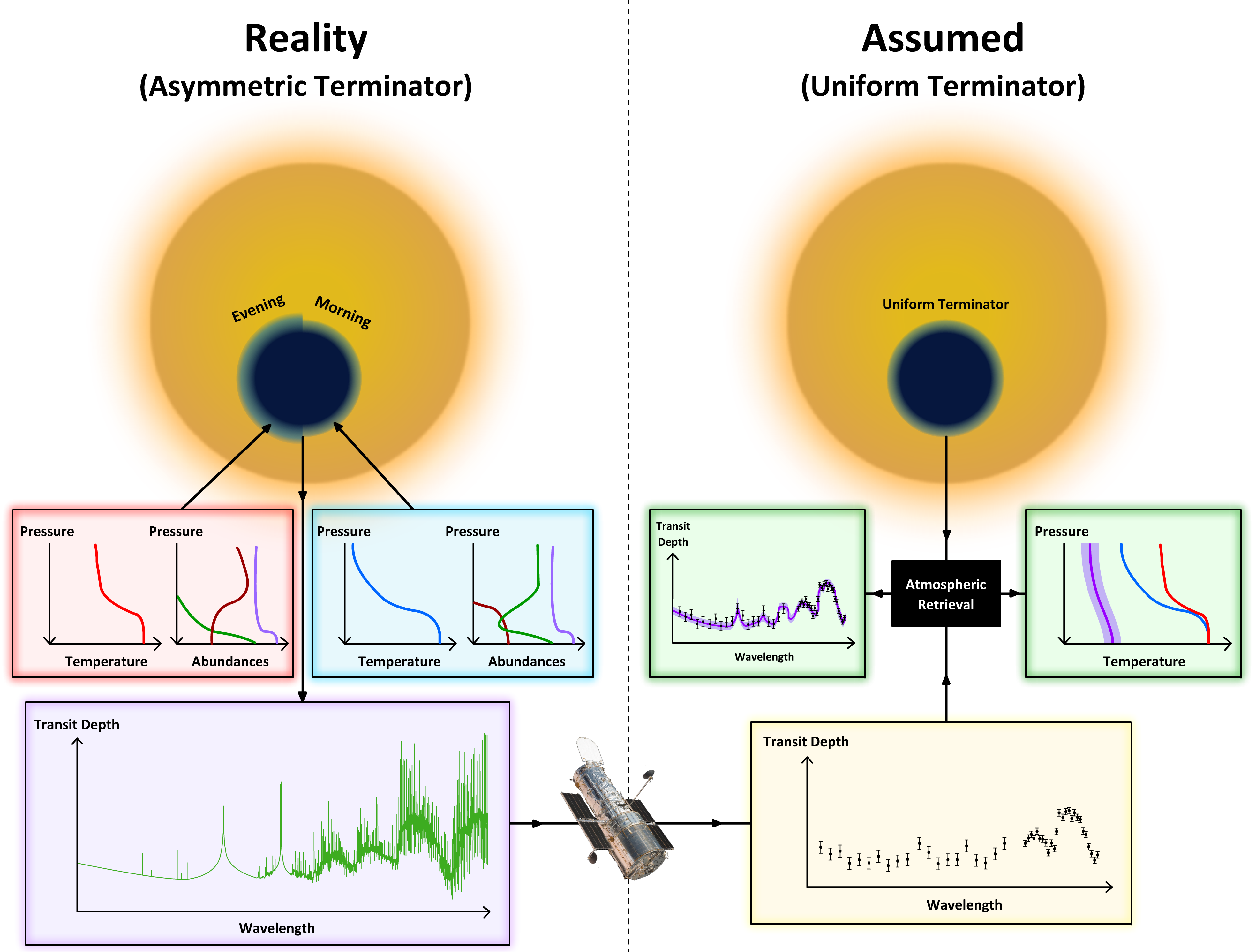}
    \caption{Schematic explanation of the cold retrieved temperatures of exoplanet terminators. Left: a transiting exoplanet with a morning-evening temperature difference (observer's perspective). Differing temperature and abundance profiles encode into the planet's transmission spectrum. Right: the observed spectrum is analysed by retrieval techniques assuming a uniform terminator. The retrieved 1D temperature profile required to fit the observations is biased to colder temperatures.}
    \label{fig:schematic_diagram}
\end{figure*}

In this study, we offer an explanation for the cold temperatures retrieved from transmission spectra. We propose these cold temperature arise from retrievals assuming 1D compositions and temperatures, such that the atmospheric properties experienced by rays traversing the terminator depend only on altitude. However, 3D General Circulation Models (GCMs) predict large gradients along the slant path (day-night differences) and azimuthally around the terminator (morning-evening differences) \citep[e.g.][]{Kataria2016,Helling2019}.

Recently, several studies have elucidated biases from 1D retrieval assumptions. \citet{Line2016} showed transmission spectra of solar-composition atmospheres with patchy clouds can be mistaken for 1D cloud-free high mean molecular weight atmospheres. \citet{Caldas2019} and \citet{Pluriel2020} found 1D transmission spectra retrievals of planets with day-night temperature and compositional gradients can be biased to higher temperatures (by $\gtrsim 200$\,K) and to erroneous abundances (with C/O overestimated by orders of magnitude). In parallel, \citet{Feng2016} and \citet{Taylor2020} showed 1D emission spectra retrievals of planets with 2D temperature structures can suffer from spurious molecular detections and abundance biases. However, these biases do not explain the cold temperatures from transmission spectra in Table~\ref{tab:litvalues}, suggesting the existence of an additional bias beyond those hitherto uncovered.

In this letter, we demonstrate that compositional differences around a terminator can bias 1D retrieved temperatures to be cooler than the true average temperature; consequently, retrieved abundances are also biased. In what follows, we explore the origin and implications of this effect. We begin with an analytic treatment, before proceeding to retrievals of synthetic exoplanet spectra.

\section{Theoretical basis: Asymmetric terminator retrieval biases} \label{sec:theory}

Consider a transiting exoplanet with a temperature difference between its morning and evening terminators, as illustrated in Figure~\ref{fig:schematic_diagram}. For tidally locked planets, this can arise from various circulation regimes between the dayside and nightside. A temperature difference can in turn induce a compositional difference -- by equilibrium or disequilibrium mechanisms -- and hence an opacity difference around the terminator. These differences are imprinted into the transmission spectrum of the planet. 

Here, we demonstrate that equating the transmission spectrum of a 2D atmosphere with a morning-evening compositional difference to a 1D atmosphere results in an erroneously cool equivalent temperature.

\newpage

\subsection{Analytic origin of 2D terminator biases} \label{subsec:analytic}

A transmission spectrum is given by the wavelength-dependent area ratio between a transiting planet and its star  
\begin{equation} \label{eq:transit_depth_general}
\Delta_{\lambda} = \frac{\pi \Rp^{2} + \displaystyle\int_{0}^{2 \pi} \displaystyle\int_{R_{\mathrm{p}}}^{\infty} \left( 1 - e^{-\tau_{\lambda}(b, \theta)} \right) \, b \, db \, d\theta}{\pi \Rstar^{2}}
\end{equation} 
where $R_{\mathrm{*}}$ and $R_{\mathrm{p}}$ are the stellar and planetary radii, respectively, $b$ is the impact parameter, $\theta$ is the azimuthal angle, and $\tau_{\lambda}(b, \theta)$ is the slant optical depth -- the extinction coefficient integrated along the line of sight. The first term represents the disc area of the opaque deep atmosphere at a reference pressure $P (r = R_{\mathrm{p}}) = P_0$. The second term gives the effective area of successive atmospheric elements in a polar coordinate system, weighted by the absorptivity of each element.

For a 1D atmosphere, this expression reduces to
\begin{equation} \label{eq:transit_depth_1D}
\Delta_{\lambda, \oneD} = \frac{\Rp^{2} + 2 \displaystyle\int_{\Rp}^{\infty} \left( 1 - e^{-\tau_{\lambda}(b)} \right) \, b \, db}{\Rstar^{2}}
\end{equation} 
Analytical tractability arises via the following assumptions: (i) constant pressure scale height with altitude (i.e. an isothermal, isocompositional, isogravitational atmosphere); (ii) hydrostatic equilibrium and the ideal gas law hold; (iii) cross sections vary weakly with pressure (i.e. $\sigma_{\lambda} (P, \, T) \approx \sigma_{\lambda} (T)$); and (iv) atmospheric altitudes satisfy $z / R_p \ll 1$. With these assumptions, it is well established that Equation~\ref{eq:transit_depth_1D} can be simply written as \citep[e.g.][]{Fortney2005,desEtangs2008,deWit2013,Betremieux2017,Heng2017}
\begin{equation} \label{eq:transit_depth_1D_analytic}
\Delta_{\lambda, \oneD} = \frac{\Rp^{2} + 2 \Rp H_{\oneD} (\gamma + \ln{\tau_{0, \lambda, \oneD}})}{\Rstar^{2}}
\end{equation} 
where $H_{\oneD} = k T_{\oneD} / \mu g$ is the scale height of the 1D atmosphere ($T_{\oneD}$, $\mu$, $g$, and $k$ being respectively the 1D temperature, mean molecular mass, surface gravity, and Boltzmann constant), $\gamma \approx 0.57722$ is the Euler-Mascheroni constant, and $\tau_{0, \lambda, \oneD}$ is given by\footnote{This form assumes extinction $\propto P$ and therefore neglects collision-induced absorption -- see \citet{deWit2013, Welbanks2019a}.}
\begin{equation} \label{eq:tau_0}
\tau_{0, \lambda, \oneD} = \frac{P_0}{k T_{\oneD}} \sqrt{2 \pi \Rp H_{\oneD}} \sum_i X_{\oneD, i} \, \sigma_{\lambda, i} (T_{\oneD})
\end{equation} 
where $X_{\oneD, i}$ and $\sigma_{\lambda, i}$ are the (1D) volume mixing ratio and absorption cross section of chemical species $i$, respectively.

For the 2D atmosphere depicted in Figure~\ref{fig:schematic_diagram}, we can write Equation~\ref{eq:transit_depth_general} as
\begin{align} \label{eq:transit_depth_2D}
\Delta_{\lambda, \twoD} = \Rstar^{-2} & \left\{ \Rp^{2} + \displaystyle\int_{\Rp}^{\infty} \left( 1 - e^{-\tau_{\lambda, \M}(b)} \right) \, b \, db \right. \nonumber \\ 
&\hspace{20pt} \left. \vphantom{\sum} + \displaystyle\int_{\Rp}^{\infty} \left( 1 - e^{-\tau_{\lambda, \E}(b)} \right) \, b \, db \right\}
\end{align}
where `E' and `M' denote the evening and morning terminators. To clearly isolate the effect of interest for the present study, we assume the atmosphere to be uniform within each terminator sector, along the day-night slant path \citep[see][]{Caldas2019}, and a morning-evening transition with negligible width (i.e. `M' and `E' both span $\Delta\theta = \pi$). Making the same simplifying assumptions as the 1D case (but now with two isotherms, $T_{\E}$ and $T_{\M}$), one obtains\footnote{This assumes the planet is spherical out to $r = R_p$, with each terminator sharing a common base pressure $P = P_0$.}
\begin{equation} \label{eq:transit_depth_2D_analytic}
\Delta_{\lambda, \twoD} = \frac{\Rp^{2} + \Rp H_{\M} (\gamma + \ln{\tau_{0, \lambda, \M}}) + \Rp H_{\E} (\gamma + \ln{\tau_{0, \lambda, \E}}) }{\Rstar^{2}}
\end{equation} 

Fitting a 1D model to a 2D spectrum is equivalent to setting $\Delta_{\lambda, \twoD} = \Delta_{\lambda, \oneD}$. One may expect this condition to result in \emph{equivalent} 1D properties given by $T_{\oneD} = \bar{T} \equiv \frac{1}{2} (T_{\E} + T_{\M})$ and $X_{\oneD, i} = \bar{X_i} \equiv \frac{1}{2} (X_{\E, i} + X_{\M, i})$ -- i.e. \emph{terminator average} temperature and mixing ratios. However, this is not the case. In Appendix~\ref{appendix_A}, we derive that the actual equivalent 1D temperature is\footnote{From here, we assume a single chemical species dominates the opacity at $\lambda$. The index $i$ is thus dropped.}

\begin{equation} \label{eq:analytic_cooling_bias}
T_{\oneD} = \bar{T} \left[ \frac{\Psi_{\lambda}}{W_{-1} \left( \Psi_{\lambda} \left(\bar{X}/X_{\oneD}\right)^2 e^{-2 (\gamma + \ln{\bar{\tau}_{0, \lambda}})}  \right) } \right] 
\end{equation} 
where
\begin{equation} \label{eq:tau_bar}
\bar{\tau}_{0, \lambda} = \frac{P_0}{k \bar{T}} \sqrt{2 \pi \Rp \bar{H}} \bar{X} \sigma_{\lambda}
\end{equation} 
\begin{equation}  \label{eq:psi}
\Psi_{\lambda} = -2 \left[ \gamma + \ln{\bar{\tau}_{0, \lambda}} + \tilde{f}\left(\frac{\Delta T} {\bar{T}}\right) + \tilde{g}\left(\frac{\Delta T}{\bar{T}}, \Delta\ln{X}\right) \right]
\end{equation}
and $W_{-1} (x)$ is the lower real branch of the Lambert W function\footnote{Defined as the inverse function of $x e^{x}$ \citep{Corless1996}.}. $\tilde{f}$ and $\tilde{g}$ are dimensionless functions of the temperature and compositional differences between the evening and morning terminators: $\Delta T \equiv \frac{1}{2} (T_{\E} - T_{\M})$, $\Delta\ln{X} \equiv \frac{1}{2} (\ln{X_{\E}} - \ln{X_{\M}})$. They are given by

\begin{align}  \label{eq:f}
\tilde{f} = - \frac{1}{4} 
& \left\{ \left(1-\frac{\Delta T}{\bar{T}}\right) \ln{\left( 1-\frac{\Delta T}{\bar{T}}\right)} \right. \nonumber \\ 
&\hspace{-2pt} \left. + \left( 1 +\frac{\Delta T}{\bar{T}}\right) \ln{\left(1+\frac{\Delta T}{\bar{T}}\right)} \right\}
\end{align}
\begin{equation}  \label{eq:g}
\tilde{g} = \left(\frac{\Delta T}{\bar{T}}\right) \Delta\ln{X} - \ln{\left[\cosh(\Delta\ln{X})\right]}
\end{equation}

We now demonstrate that Equation~\ref{eq:analytic_cooling_bias} predicts $T_{\oneD} < \bar{T}$ readily occurs for $\Delta\ln{X} \neq 0$ when $X_{\oneD} = \bar{X}$ (i.e. assuming mixing ratios are correctly retrieved -- we revisit this in section~\ref{subsec:assumptions}).

\subsection{Properties of the analytic solution} \label{subsec:analytic_properties}

\begin{figure}[ht!]
    \centering
    \includegraphics[width=\columnwidth]{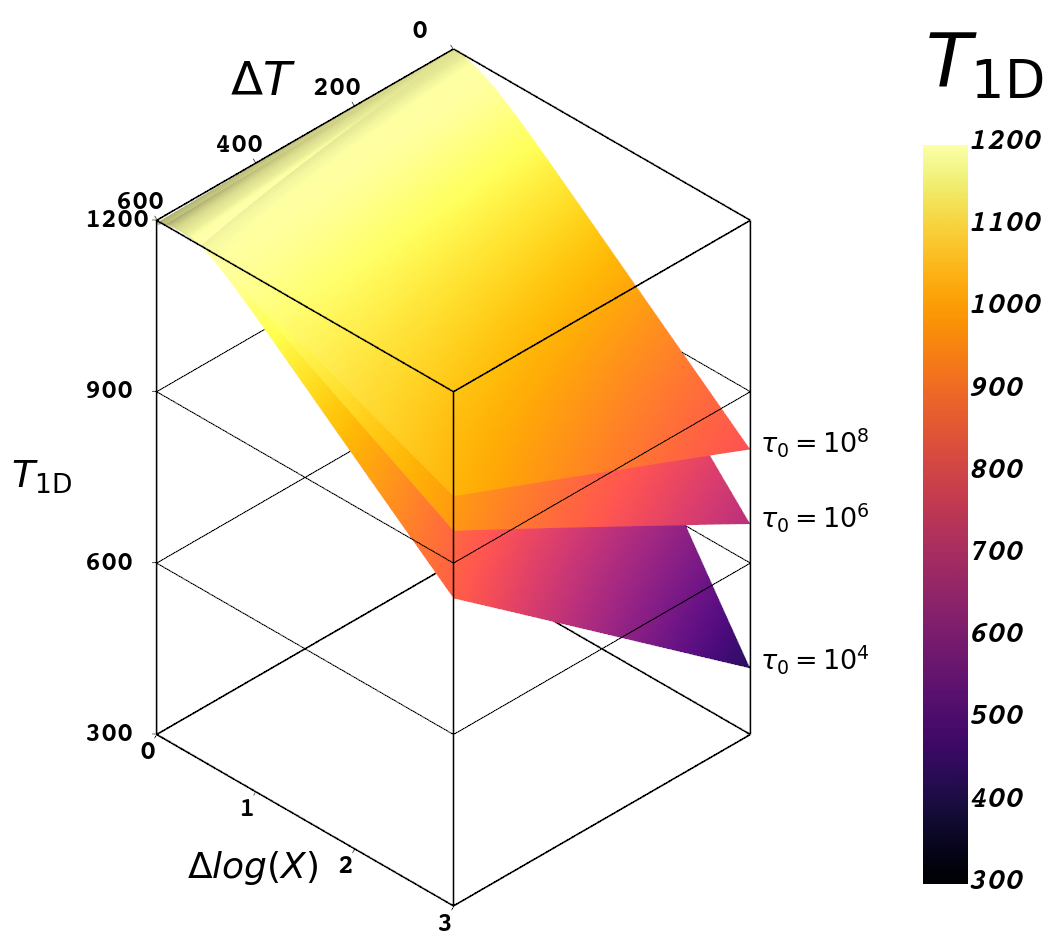}
    \caption{Analytic exploration of 2D terminator temperature biases. The temperature of a 1D atmosphere (colourbar) with an equivalent transit depth to a 2D atmosphere with terminator temperature difference $\Delta T$ and compositional difference $\Delta \mathrm{log_{10}} (X)$. Surfaces are plotted for $\bar{T} = 1200$\,K with different mean reference optical depths, $\bar{\tau}_{0, \lambda}$, according to Equation~\ref{eq:analytic_cooling_bias}. A substantial bias to cold temperatures $T_{\oneD} < \bar{T}$ arises from the influence of compositional differences. An animated version of this figure, showing a $360\degr$ rotation, is available in the HTML version of this article.}
    \label{fig:analytic_solution}
\end{figure}

Equation~\ref{eq:analytic_cooling_bias} is graphically rendered in Figure~\ref{fig:analytic_solution} for a typical hot Jupiter ($\bar{T} = 1200$\,K) over a range of $\bar{\tau}_{0, \lambda}$. Specifically, we show three $\bar{\tau}_{0, \lambda}$ surfaces considering different potential opacity sources for a hot Jupiter with $R_{\rm p} = 1.4 R_{\rm J}$, $g = 10$\,ms$^{-2}$, and $\mu = 2.3\,m_{\rm u}$. The first surface ($\bar{\tau}_{0, \lambda} = 10^4$) corresponds to H$_2$O absorption around $1.4\,\micron$ ($\sigma_{\lambda} \sim 10^{-21}$\,cm$^{2}$, \citet{Sharp2007}) with $X_{\rm{H_2 O}} = 10^{-4}$. The second surface ($\bar{\tau}_{0, \lambda} = 10^6$) corresponds to TiO absorption around $0.7\,\micron$ ($\sigma_{\lambda} \sim 10^{-16}$\,cm$^{2}$, \citet{Sharp2007}) with $X_{\rm{TiO}} = 10^{-7}$. Finally, the third surface ($\bar{\tau}_{0, \lambda} = 10^8$) corresponds to Na doublet line core absorption around $0.6\,\micron$ ($\sigma_{\lambda} \sim 10^{-15}$\,cm$^{2}$, \citet{Allard2019}) with $X_{\rm{Na}} = 10^{-6}$. From examining Figure~\ref{fig:analytic_solution}, one notices three important takeaways:

\begin{enumerate}
    \item Pure temperature differences ($\Delta\ln{X} = 0$) result in negligible biases to retrieved temperatures.
    \item Compositional differences exceeding a factor 2 ($\Delta \log_{10} (X) > 0.3$) result in $T_{\oneD}$ biases to many hundreds of degrees colder than $\bar{T}$.
    \item The wavelength dependency of $T_{\oneD}$ (from $\bar{\tau}_{0, \lambda}$) implies that no one equivalent temperature can perfectly reproduce a 2D spectrum using a 1D model.
\end{enumerate}

Observations 1 and 2 arise from properties of $\tilde{f}$ and $\tilde{g}$, elucidated in Appendix~\ref{appendix_B}. The final observation yields an important consequence for fitting 2D transmission spectra with 1D models: as a retrieval can only chose one value of $T_{\oneD} \, \forall \, \lambda$, the chosen value will balance the different $\bar{\tau}_{0, \lambda}$ surfaces to minimise residuals between the 2D data and 1D model. The chosen $T_{\oneD}$, considered as a wavelength-average, will however still be biased to colder temperatures than $\bar{T}$ (Figure~\ref{fig:analytic_solution}). It may be possible to use such residuals, with sufficiently precise observations, to identify a preference for 2D atmospheric models. 

\subsection{Limiting assumptions} \label{subsec:assumptions}

The conceptual picture provided by Equation~\ref{eq:analytic_cooling_bias} and Figure~\ref{fig:analytic_solution} will be altered in regimes where the underlying assumptions break down. We highlight two important limitations: (i) if $T_{\oneD}$ becomes sufficiently cooler than $\bar{T}$, neglected temperature dependencies in absorption cross sections, $\sigma_{\lambda} (T)$, will alter the shapes of spectral features; and (ii) as $\Delta\ln{X}$ grows, the assumption that only one species dominates the opacity will be violated on the terminator side deficient in the given species. Taken together, the breakdown of these assumptions will place a lower limit on how cold $T_{\oneD}$ may become.

1D models have two additional degrees of freedom to compensate for such higher order effects: $X_{\oneD}$ and $\Rp$. Varying the former is already encapsulated by Equation~\ref{eq:analytic_cooling_bias}. Varying the latter corresponds to the retrieved base planet radius, $\RponeD$, differing from the actual base radius, $\RptwoD$ (equated until now). A generalisation of Equation~\ref{eq:analytic_cooling_bias} for $\RponeD \neq \RptwoD$ is presented in Appendix~\ref{appendix_A}. Perturbing either $X_{\oneD}$ or $R_{p, \oneD}$ (therefore biasing these quantities) effectively translates the surfaces in Figure~\ref{fig:analytic_solution} along the $T_{\oneD}$ axis (not shown). We thus expect that some of the $T_{\oneD}$ bias will be `shifted' into $X_{\oneD}$ and $\RponeD$, each attaining their own bias. 

To relax many of the aforementioned assumptions, and establish the extent of $T_{\oneD}$, $X_{\oneD}$, and $\RponeD$ biases, we turn to more physically realistic numerical models. 

\section{Exploration of 2D retrieval biases} \label{sec:retrievals_toy_planets}

We explore here the degree to which asymmetric terminators can confound 1D atmospheric retrieval techniques. Our strategy follows a four-step approach: (i) generate model transmission spectra for range of atmospheres with asymmetric (2D) terminators; (ii) convolve the 2D models to a spectral resolution and precision typical of current HST observations; (iii) run the synthetic data through a 1D retrieval code; and (iv) compare the retrieved 1D atmospheric properties to the true terminator-averaged properties. In turn, we describe our atmospheric case studies, the modelling and retrieval procedure, and present the resulting biases for each case.

\subsection{Atmospheric case studies} \label{subsec:case_studies}

As the cold retrieved temperatures of exoplanets span a wide range of equilibrium temperatures (Table~\ref{tab:litvalues}), so too must our proposed explanation. We therefore consider three diverse case studies of atmospheres expected to posses morning-evening compositional differences:

\begin{enumerate}
    \item \textbf{Warm Jupiter}: $\bar{T}_{\rm{1 \, mbar}} \sim 1000$\,K, $\Delta T = 100$\,K (i.e. a $200$\,K morning-evening temperature difference). The warmer (evening) terminator has Na and K abundances representative of solar elemental abundances: log($X_{\E, \, \rm{Na}}$) = -6, log($X_{\E, \, \rm{K}}$) = -7 \citep{Asplund2009}. The cooler (morning) terminator is assumed depleted in Na and K by 2 orders of magnitude (a proxy for condensation). CH$_4$ roughly follows equilibrium abundances for a solar-composition atmosphere: log($X_{\E, \, \rm{CH_4}}$) = -6, log($X_{\M, \, \rm{CH_4}}$) = -4 \citep{Heng2016}. H$_2$O takes a solar abundance, assumed constant around the terminator: log($X_{\E, \, \rm{H_2 O}}$) = log($X_{\M, \, \rm{H_2 O}}$) = -3.3.
    
    \item \textbf{Hot Jupiter}: $\bar{T}_{\rm{1 \, mbar}} \sim 1600$\,K, $\Delta T = 100$\,K. Both terminators possess constant Na, K, and H$_2$O abundances: log($X_{\rm{Na}}$) = -6, log($X_{\rm{K}}$) = -7, log($X_{\rm{H_2 O}}$) = -3.3. The evening terminator additionally contains TiO and VO with roughly solar abundances: log($X_{\E, \, \rm{TiO}}$) = -7, log($X_{\E, \, \rm{VO}}$) = -8. The morning terminator is assumed sufficiently cool for all TiO and VO to have condensed out of the gas-phase in the observable atmosphere.
    
    \item \textbf{Ultra-hot Jupiter}: $\bar{T}_{\rm{1 \, mbar}} \sim 2200$\,K, $\Delta T = 250$\,K. Both terminators possess Na, K, and H$_2$O abundances as in the hot Jupiter case. However, here the evening terminator is warm enough for H$_2$ to partially dissociate and form an inventory of the hydrogen anion \citep{Parmentier2018}, for which we take: log($X_{\E, \, \rm{H^{-}}}$) = -8. The morning terminator is assumed too cold to support H$^{-}$.
\end{enumerate}

In all three cases, physical properties are representative of HD~209458b: $R_{\rm p} = 1.359 R_{\rm J}$\footnote{The reference radius is set to 98\% of the white light radius: $R_{\rm{p, \, 10 \, bar}} = 1.33182 R_{\rm J}$.}, $M_{\rm p} = 0.6845 M_{\rm J}$. Each planet is assumed H$_2$+He dominated, with a solar-proportion He/H$_2$ ratio of 0.17. The pressure-temperature (P-T) profiles and morning-evening temperature differences are inspired by literature GCM profiles \citep[e.g.][]{Kataria2016,Helling2019}, constructed parametrically \citep{Madhusudhan2009}\footnote{The warm and hot Jupiters have $\alpha_{1, \, [M, E]}$ = [0.6, 0.7], $\alpha_{2, \, [M, E]}$ = [0.5, 0.6], log($P_{1}$) = -2.0, log($P_{2}$) = -5.0, and log($P_{3}$) = 1.0. The ultra-hot Jupiter instead has $\alpha_{1, \, [M, E]}$ = [0.5, 0.7], $\alpha_{2, \, [M, E]}$ = [0.4, 0.6], with the pressure parameters as previous.}. Each profile has an `anchor' temperature at 10~bar, $T_{\rm deep}$, below which the atmosphere is homogeneous: $T_{\rm deep, \, warm} = 1600$\,K, $T_{\rm deep, \, hot} = 2200$\,K, $T_{\rm deep, \, ultra-hot} = 3000$\,K. To isolate biases arising solely from the compositional and temperature differences, the terminators are assumed cloud-free. We note that the morning-evening differences assumed here are intended as illustrative of their corresponding biases, with strictly self-consistent profiles considered in section~\ref{sec:real_planets}.   

\subsection{Modelling \& retrieval procedure} \label{subsec:sec3_methods}

\begin{figure*}[ht!]
    \centering
    \includegraphics[width=0.94\textwidth]{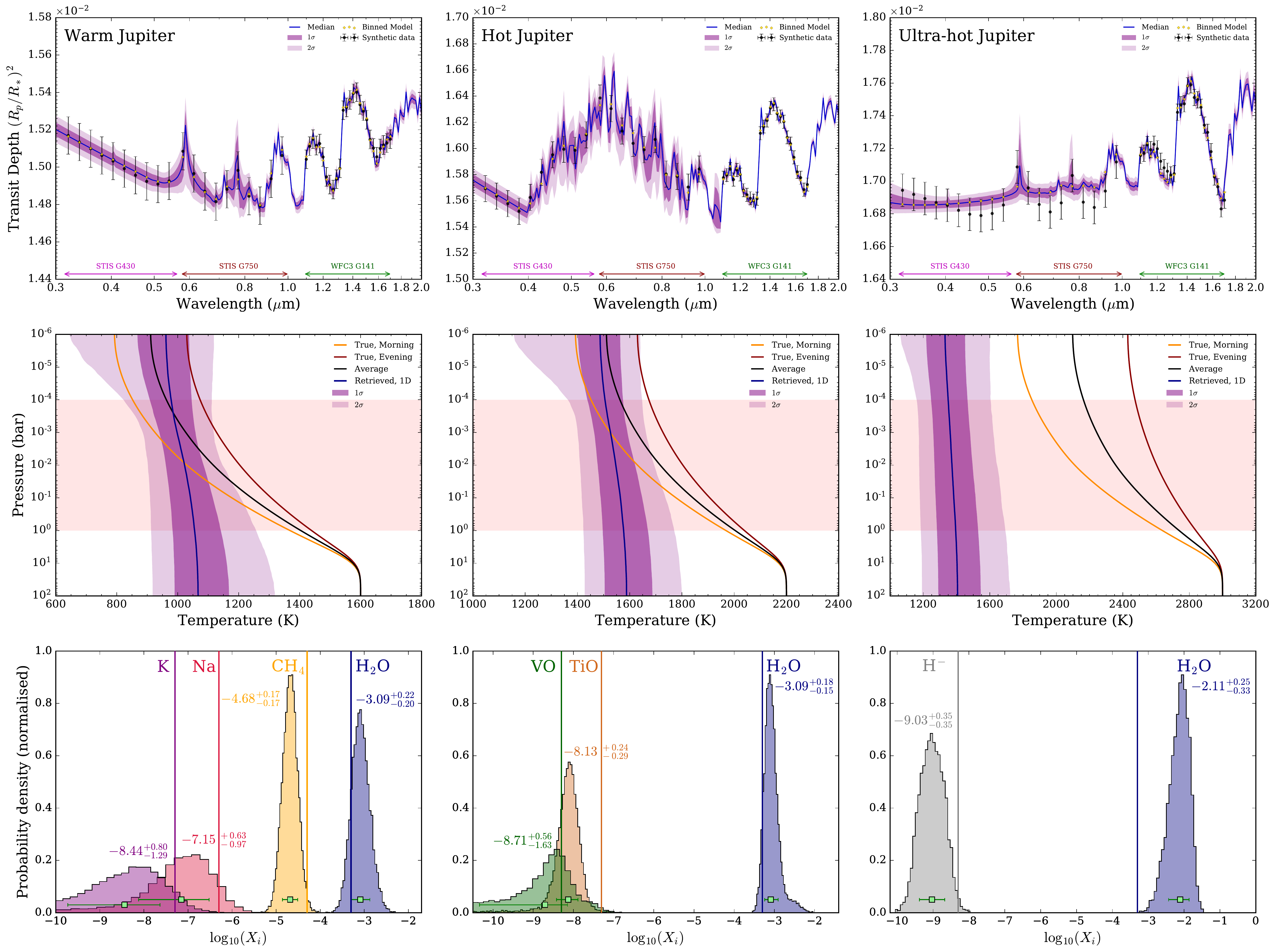}
    \caption{Numerical exploration of 2D terminator retrieval biases. Each column covers a different atmospheric case study: a $\sim 1000$\,K warm Jupiter, $\sim 1600$\,K hot Jupiter, and $\sim 2200$\,K ultra-hot Jupiter (see text for details). Top row: 2D model transmission spectra binned to typical HST STIS and WFC3 spectral resolutions ($R = 20$, $60$) and precisions ($100$\,ppm, $50$\,ppm). A 1D retrieval of this data yields the coloured confidence regions. Middle row: true morning (orange) and evening (red) P-T profiles used to generate each 2D model, alongside the terminator-averaged profile (black) and retrieved 1D profile (coloured contours). The pressure range typically probed by the spectra, $10^{-4} - 1\,$bar, is shaded red. Bottom row: retrieved 1D mixing ratio posteriors. The true terminator-averaged abundances (solid lines) are compared to the retrieved 1D abundances (labels). Na and K are omitted from the hot and ultra-hot Jupiter posteriors as they are relatively unconstrained (see the \href{https://www.doi.org/10.5281/zenodo.3723448}{online posteriors}). The retrieved 1D P-T profiles are biased to colder temperatures, while the retrieved abundances can be biased in either direction.}
    \label{fig:retrieval_multi_panel}
\end{figure*}

Our model atmospheres, radiative transfer, and retrievals are computed using the POSEIDON atmospheric retrieval code \citep{MacDonald2017}. The atmospheric column in each terminator is discretised uniformly in log-pressure with 10 layers per decade, for 81 levels from 10$^{-6}$ to 10$^2$ bar. The deep atmosphere ($P \geq 10$\, bar) has a homogeneous temperature, by construction, hence the terminators share a spherical radial grid below the reference radius $R_{\rm{p, \, 10 \, bar}}$. Above this, separate radial grids are constructed for each terminator using P-T profiles, abundances, and planetary properties under hydrostatic equilibrium. The warmer evening terminator hence extends to greater radii (see Figure~\ref{fig:schematic_diagram}). The opacities of Na, K, H$_2$O, CH$_4$, TiO, and VO, cross section computations, broadening parameters, and line list references are described in \citet{MacDonald_Thesis_2019}. H$^{-}$ continuum opacity is included \citep{John1988}. High-resolution ($R \approx 10^6$) opacities are sampled onto a $R = 2000$ wavelength grid from $0.3 - 2.0\,\micron$. Radiative transfer is solved separately for each terminator, with 2D transmission spectra constructed by a linear superposition of each terminator spectrum.  

Synthetic Hubble observations are generated by convolving each model spectrum to the resolving power of the STIS G430 / G750 and WFC3 G141 grisms, before integrating over their respective sensitivity functions. We choose spectral resolutions and precisions typical of current HST observations \citep[e.g.][]{Sing2016}: $R = 20$ and 100 ppm for STIS \& $R = 60$ and 50 ppm for WFC3. The synthetic data are placed on the true transit depths (i.e. without Gaussian scatter), such that any posterior deviations from the true parameter values are attributable to a retrieval bias rather than a specific noise instance \citep[see][]{Feng2018}. The resulting data are shown in Figure~\ref{fig:retrieval_multi_panel} (top row).

\newpage

We subject each synthetic dataset to a Bayesian atmospheric retrieval. The retrievals assume a 1D forward model with a single 6 parameter P-T profile \citep{Madhusudhan2009}, a single abundance for each species, and a 10 bar planetary radius. The warm and ultra-hot Jupiter retrievals have 11 free parameters, whilst the hot Jupiter retrieval has 12. The P-T parameter priors are as described in \citet{MacDonald2019}, with $T_{\rm{deep}}$ ascribed a uniform prior from $400 - 3000\,$K. The logarithm of each mixing ratio has a uniform prior from -12 to -0.3. $R_{\rm{p, \, 10 \, bar}}$ has a uniform prior from $0.85 - 1.15\,\Rp$. The parameter space is explored via the nested sampling algorithm MultiNest \citep{Feroz2008,Feroz2009,Feroz2013}, implemented by the python wrapper PyMultiNest \citep{Buchner2014}.

\subsection{Results: retrieval biases} \label{subsec:sec3_results}

Our retrieved spectra, P-T profiles, and mixing ratios are shown\footnote{Posteriors are available at: \href{https://www.doi.org/10.5281/zenodo.3723448}{doi.org/10.5281/zenodo.3723448}.} in Figure~\ref{fig:retrieval_multi_panel}. The 1D models achieve an excellent fit for the warm and hot Jupiters, with the median models lying within $1\sigma$ of all data points (i.e. discrepancies $<$ 50 ppm). The ultra-hot Jupiter spectral fit is the least accurate, with around 15\% of the data incorrectly fit to $1\sigma$. The latter observation, most prominent at visible wavelengths, arises from a 1D model dominated by H$^{-}$ being unable to reproduce the superposition of H$^{-}$ and H$_2$ Rayleigh scattering encoded in the 2D model. However, for real observations with Gaussian scatter, it would be difficult to recognise such a model mismatch.

The retrieved P-T profiles are colder than the terminator-averaged profiles for $P >\,$1 mbar. At a 10 mbar reference level ($\sim$ the median photosphere), biases of -100\,K, -200\,K, and -1000\,K result for the warm, hot, and ultra-hot cases, respectively. The retrieved profiles exhibit shallower temperature gradients than the true profiles, despite the ability of 1D retrievals to retrieve temperature gradients \citep{Rocchetto2016,MacDonald2017}, possibly explaining why many retrieved P-T profiles appear near-isothermal \citep[e.g.][]{Pinhas2019,MacDonald2019}. Our findings match the general trend seen in Table~\ref{tab:litvalues}: retrieved temperatures become far colder than expectations (cf. skin temperatures) as $T_{\rm{eq}}$ rises.

The mixing ratios for all significant opacity sources\footnote{Only VO (hot Jupiter case), Na and K (hot + ultra-hot cases) are retrieved within $1\sigma$. VO is almost obscured by TiO, whilst the alkalis are unconstrained in the hot + ultra-hot cases.} are incorrectly retrieved to $1\sigma$. Many chemical species are only retrieved accurately to the $3\sigma$ level (e.g. CH$_4$, TiO, and H$^{-}$). Those species exhibiting compositional differences have retrieved 1D abundances biased lower than the true terminator-averaged values. Even species uniform around the terminator (here, H$_2$O) are biased, though to higher abundances. Compositional biases become more severe as the retrieved P-T profile deviates further from the true terminator temperature. In the most extreme case, the retrieved H$_2$O abundance is biased by over an order of magnitude, such that one would incorrectly believe a solar-metallicity atmosphere was $15\,\times$ super-solar at $> 3\sigma$ confidence. 

\section{Application to specific planets} \label{sec:real_planets}

Finally, we demonstrate that asymmetric terminators can naturally explain the cold retrieved temperatures of specific exoplanet atmospheres. We consider self-consistently calculated temperature structures and compositions for one hot Jupiter (WASP-17b) and one ultra-hot Jupiter (WASP-12b). This allows the extension of section~\ref{sec:retrievals_toy_planets}'s results to consider changing compositions with altitude, due to effects such as TiO condensation and H$_2$O dissociation, in a self-consistent manner.

\subsection{Self-consistent atmospheric models} \label{subsec:self_consistent}

We compute self-consistent atmospheric P-T and compositional profiles under the assumptions of radiative-convective and chemical equilibrium with rainout condensation and ionisation. An initial P-T profile is iteratively perturbed until a solution satisfying hydrostatic equilibrium, energy conservation, and equilibrium chemistry is obtained. Our models include all the opacities from \citet{Goyal2018}, alongside Fe and H$^{-}$. The model is fully described in \citet{Goyal_Thesis_2019} and Goyal et al. (in prep). We simulate P-T profiles for each terminator by varying the recirculation factor \citep[e.g.][]{Fortney2007} - a 1D proxy for advection due to winds.

Our self-consistent atmospheric profiles are shown in Figure~\ref{fig:retrieval_real_planets}. WASP-17b displays large compositional differences in TiO and VO for $P \lesssim 10^{-2}$\,bar due to metal oxide condensation on the cooler morning terminator. We therefore add an opaque cloud deck to the atmospheric model at 10 mbar, serving as a proxy for Ti and V condensates. At the higher temperature of WASP-12b, compositional differences instead arise from H$_2$ and H$_2$O dissociation in the upper atmosphere.  

Transmission spectra are computed for each planet following the same methodology as section~\ref{subsec:sec3_methods}. Opacity and mean molecular weight contributions from the altitude-dependant abundances of H$_2$, He, H, H$^{-}$, Na, K, H$_2$O, TiO, and VO are considered. For WASP-12b we discount TiO and VO, as their large spectral signatures are inconsistent with current observations at optical wavelengths \citep{Kreidberg2015,Sing2016}. The resultant 2D spectra are convolved to a similar resolution and precision as literature spectra for each planet \citep{Sing2016}: 200~ppm / 400~ppm for WASP-17b (STIS / WFC3) and 100~ppm / 50~ppm for WASP-12b. Retrievals are conducted as previously, with the addition of a cloud pressure parameter, $P_{\rm{cloud}}$, for WASP-17b.

\subsection{Retrieval biases: WASP-17b \& WASP-12b} \label{subsec:sec4_results}

\begin{figure*}[ht!]
    \centering
    \includegraphics[width=\textwidth]{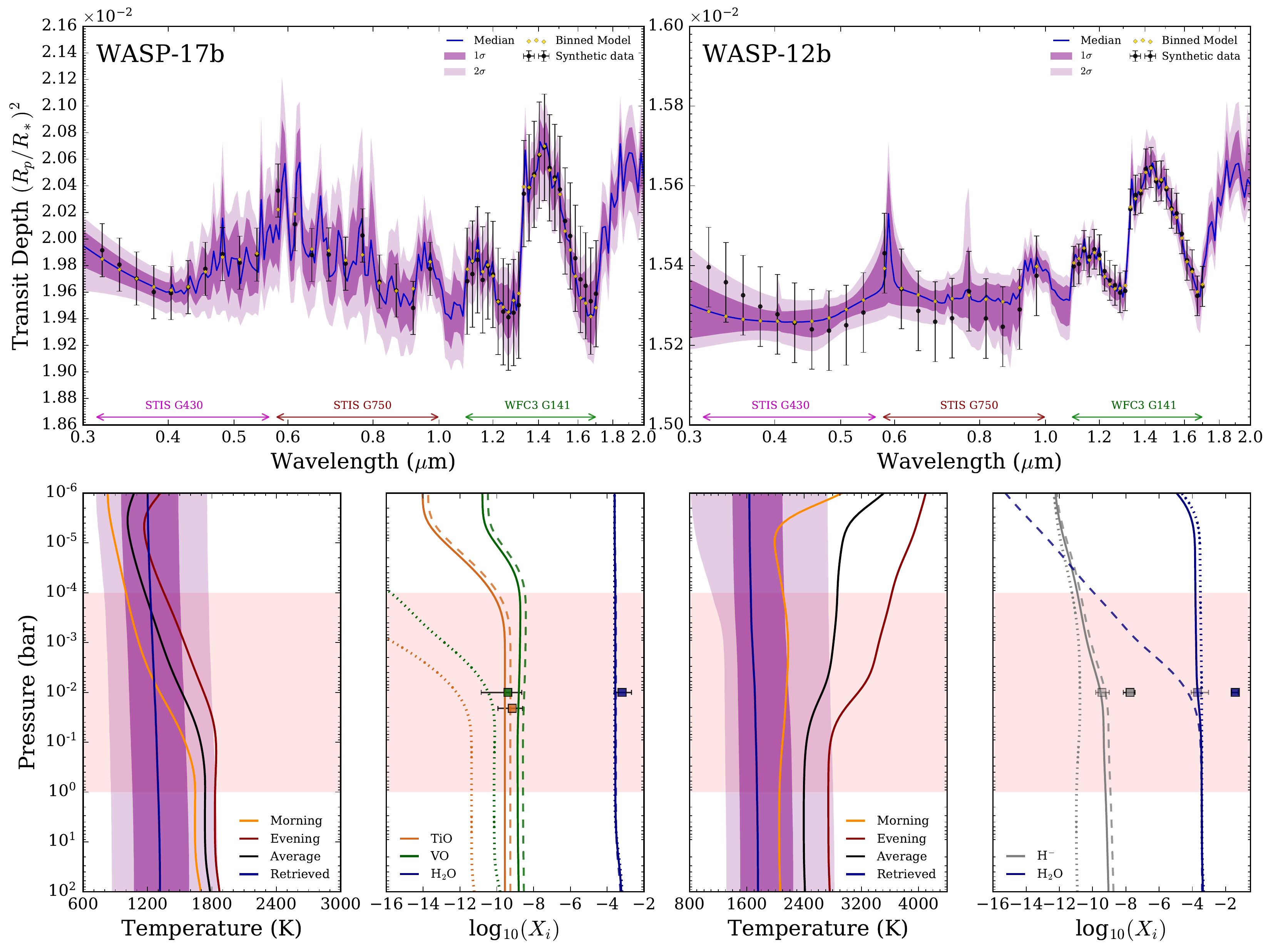}
    \caption{Influence of 2D atmosphere retrieval biases on self-consistent exoplanet atmospheres. Top row: 2D model transmission spectra for WASP-17b (left) and WASP-12b (right), forming synthetic HST STIS and WFC3 data with similar precisions to each planet's observed spectra \citep{Sing2016}. A 1D retrieval yields the coloured confidence regions. Lower left: true morning (orange) and evening (red) P-T profiles, alongside the terminator-averaged profile (black) and retrieved 1D profile (coloured contours). The pressure range typically probed, $10^{-4} - 1\,$bar, is shaded red. Lower right: mixing ratio profiles for the morning (dotted) and evening (dashed) terminators. The true terminator-averaged abundances (solid) are compared to the retrieved 1D abundances (error bars). WASP-12b has a bimodal abundance solution, with the sub-dominant mode rendered by transparent error bars. Biases are greatest for WASP-12b, with a cooler P-T profile and erroneously enhanced abundances.}
    \label{fig:retrieval_real_planets}
\end{figure*}

Retrieval results for each planet are shown in Figure~\ref{fig:retrieval_real_planets}. We obtain biased retrieved P-T profiles and abundances, despite 1D model spectra providing adequate fits to the 2D atmosphere spectra. Both planets have retrieved P-T profiles biased to colder temperatures, with more extreme biases for the ultra-hot Jupiter WASP-12b.  

WASP-17b provides an example where retrieval biases are reasonably contained. Our retrieved photosphere temperature, $T_{\rm{10 \, mbar}} = 1265^{+273}_{-237}$\,K, is biased by $\sim 250$\,K below the true terminator-averaged temperature. This is consistent with reported literature temperatures \citep{Pinhas2019,Welbanks2019b}. The retrieved abundances of TiO, VO, and H$_2$O agree with the true terminator-averaged values to $1\sigma$. This demonstrates that abundances derived from transmission spectra with large error bars ($> 100$\,ppm) can still be considered accurate despite 2D terminator differences. 

WASP-12b provides a cautionary tale for how 1D retrievals can infer erroneous atmospheric properties. The retrieved temperature, $T_{\rm{10 \, mbar}} = 1711^{+472}_{-236}$\,K, is biased by almost 1000\,K below the true terminator-averaged temperature. This is consistent with the cold temperature derived by \citet{Kreidberg2015} to $1\sigma$ and that of \citet{Pinhas2019} and \citet{Welbanks2019b} to $2\sigma$ (Table~\ref{tab:litvalues}). The abundance posteriors display a bimodal solution (see the \href{https://www.doi.org/10.5281/zenodo.3723448}{online posteriors}), with a high-metallicity mode (maximum likelihood) and a roughly solar metallicity mode. We represent each solution separately in Figure~\ref{fig:retrieval_real_planets}, showing two $1\sigma$ error bars derived from separate retrievals with a cut imposed at log($X_{\rm{H_2 O}}$) = -2.  The preferred mode has H$^{-}$ and H$_2$O abundances biased by nearly 2 orders of magnitude: log($X_{\rm{H^{-}}}$) = $-7.75^{+0.31}_{-0.41}$ and log($X_{\rm{H_2 O}}$) = $-1.41^{+0.17}_{-0.21}$, echoing the lesson of section~\ref{sec:retrievals_toy_planets} that 1D abundances for ultra-hot Jupiters must be carefully considered.

\newpage

\section{Summary \& Discussion}
\label{sec:Discussion}

The retrieved cold temperatures of exoplanet terminators in the literature can be explained by inhomogenous morning-evening terminator compositions. The inferred temperatures arise from retrievals assuming uniform terminator properties. We have demonstrated analytically that the transit depth of a planet with different morning and evening terminator compositions, when equated to a 1D transit depth, results in a substantially colder temperature than the true average terminator temperature. This also holds for state-of-the-art retrieval codes, with the added complication that retrieved chemical abundances can also be significantly biased.

Our results have several implications for atmospheric studies of exoplanets:

\begin{enumerate}
    \item Transmission spectra of planets with asymmetric terminators can be fit by 1D models, but the inferred atmospheric properties may not represent their true terminator-averages.
    \item The temperatures of exoplanet terminators reported in the literature may be biased by several hundred degrees below their true value. The biases are most extreme for ultra-hot Jupiters, reaching $\sim 1000$\,K. Retrieved 1D temperature structures of asymmetric terminators also exhibit much weaker temperature gradients than those really present.
    \item Chemical abundances derived from 1D retrieval techniques are often biased by greater than the quoted $1\sigma$ uncertainty, even if the species in question is uniform around the terminator. For ultra-hot Jupiters, such biases can exceed $3\sigma$. These biases may limit our ability to robustly constrain planetary formation mechanisms from retrieved atmospheric compositions.
\end{enumerate}

Our study has also revealed that 1D models do not provide perfect fits at all wavelengths to transmission spectra of planets with asymmetric terminators. Residuals present in 1D model fits offer the promise that sufficiently precise observations would yield a clear preference for 2D atmospheric models. Retrieval techniques with more sophisticated forward models may therefore be able to exploit these residuals to correctly infer unbiased properties of exoplanet terminators.

There is an urgent need to extend retrieval techniques to account for non-uniform atmospheres. Our findings add to the growing evidence that transmission spectra encode far more information than we can access by the application of 1D models \citep{Fortney2010,Line2016,Caldas2019,Pluriel2020}. Developing retrieval tools capable of accounting for 2D or 3D atmospheric properties will necessitate larger parameter spaces and increased computational burdens. Nevertheless, the opportunities afforded by such endeavours are immense, offering a rich multidimensional window into the atmospheres of these distant worlds.

\vspace{-0.2cm}

\acknowledgments
\noindent We extend gratitude to the anonymous referee, whose insightful comments improved the quality of this paper.

\appendix
\section{Derivation of the 1D temperature equivalent to a 2D transmission spectrum} \label{appendix_A}

We wish to find the temperature a 1D model must take to produce the same transit depth as a 2D atmosphere with differing morning and evening terminators (Figure~\ref{fig:schematic_diagram}). Following standard assumptions rendering analytic tractability to the transit depth integral in Equation~\ref{eq:transit_depth_1D}, the condition $\Delta_{\lambda, \twoD} = \Delta_{\lambda, \oneD}$ can be written as
\begin{equation} \label{eq:A1}
\frac{\RponeD^{2} + 2 \RponeD H_{\oneD} (\gamma + \ln{\tau_{0, \lambda, \oneD}})}{\Rstar^{2}} = \frac{\RptwoD^{2} + \RptwoD H_{\M} (\gamma + \ln{\tau_{0, \lambda, \M}}) + \RptwoD H_{\E} (\gamma + \ln{\tau_{0, \lambda, \E}}) }{\Rstar^{2}}
\end{equation} 
where 
\begin{align} \label{eq:A2_A3}
\tau_{0, \lambda, \oneD} &= \frac{P_0}{k T_{\oneD}} \sqrt{2 \pi \RponeD H_{\oneD}} \sum_i X_{\oneD, i} \, \sigma_{\lambda, i} (T_{\oneD}) \\
\tau_{0, \lambda, (\M / \E)} &= \frac{P_0}{k T_{(\M / \E)}} \sqrt{2 \pi \RptwoD H_{(\M / \E)}} \sum_i X_{(\M / \E), i} \, \sigma_{\lambda, i} (T_{(\M / \E)})
\end{align} 
This general form assumes a common base pressure, $P_0$, for the morning and evening terminators. Due to the differing temperatures in each terminator, the radius corresponding to a given pressure diverges for altitudes above this reference level. $\RponeD$ and $\RptwoD$ are defined according to $r_{\oneD} (P = P_0) = \RponeD$ and $r_{\twoD, \, (\M / \E)} (P = P_0) = \RptwoD$. As the atmosphere is assumed opaque below $\RponeD$ and $\RptwoD$ in deriving Equations~\ref{eq:transit_depth_1D_analytic} and \ref{eq:transit_depth_2D_analytic}, $P_0$ must be sufficiently deep to satisfy $\tau_{\lambda} \gg 1 \, \forall \lambda$ ($P_0$ is hence often taken as 10 bar in retrieval studies).

In what follows, two key assumptions will be made: 

\begin{enumerate}
    \item $\sigma_{\lambda, i} (T)$ varies sufficiently weakly between the terminators that the temperature dependence can be dropped. This amounts to a zeroth order Taylor expansion about a reference temperature, which we take as $\bar{T} \equiv \frac{1}{2} (T_{\E} + T_{\M})$. Note the caveat in section~\ref{subsec:assumptions} that this will break down for $T_{\oneD} \ll \bar{T}$.
    \item The extinction at a given wavelength $\lambda$ is dominated by a single chemical species, such that $\sum_i X_{i} \sigma_{\lambda, i} \approx X \sigma_{\lambda}$.
\end{enumerate}
For notational convenience, we hence drop the species index `$i$' and cross section temperature dependence. Given these assumptions and notational conventions, Equation~\ref{eq:A1} can be rearranged to
\begin{align} \label{eq:A4}
& 2 H_{\oneD} \left[ \gamma + \ln{\left( \frac{P_0}{k T_{\oneD}} \sqrt{2 \pi \RponeD H_{\oneD}} \, X_{\oneD} \, \sigma_{\lambda} \right)} \right] = \frac{\RptwoD^{2} - \RponeD^{2}}{\RponeD} \, + \nonumber \\  \frac{\RptwoD}{\RponeD} & \left\{ H_{\M} \left[ \gamma + \ln{\left( \frac{P_0}{k T_{\M}} \sqrt{2 \pi \RptwoD H_{\M}} \, X_{\M} \, \sigma_{\lambda} \right)} \right] + H_{\E} \left[ \gamma + \ln{\left( \frac{P_0}{k T_{\E}} \sqrt{2 \pi \RptwoD H_{\E}} \, X_{\E} \, \sigma_{\lambda} \right)} \right] \right\}
\end{align} 

Our goal is to solve for $T_{\oneD}$ (or equivalently, $H_{\oneD}$). We first rewrite the LHS of Equation~\ref{eq:A4} by inserting four factors of unity ($1 = \bar{T}/\bar{T} = \bar{H}/\bar{H} = \bar{X}/\bar{X} = \RptwoD/\RptwoD$), such that
\begin{equation} \label{eq:A5}
\mathrm{LHS} = 2 H_{\oneD} \left[ \gamma + \ln{\bar{\tau}_{0, \lambda}} + \ln{\sqrt{\frac{\bar{H}}{H_{\oneD}}}} + \ln{\left(\frac{X_{\oneD}}{\bar{X}}\right)} + \ln{\sqrt{\frac{\RponeD}{\RptwoD}}} \right]
\end{equation} 
where
\begin{equation} \label{eq:A6}
\bar{\tau}_{0, \lambda} = \frac{P_0}{k \bar{T}} \sqrt{2 \pi \RptwoD \bar{H}} \bar{X} \sigma_{\lambda} = \frac{P_0}{\mu g} \sqrt{\frac{2 \pi \RptwoD}{\bar{H}}} \bar{X} \sigma_{\lambda}
\end{equation} 
and the relation $H = k T / \mu g$ has been used to encode all temperature dependencies in terms of scale heights. 

Considering that the null hypothesis of unbiased transmission spectra would yield $H_{\oneD} = \bar{H} \equiv \frac{1}{2} (H_{\E} + H_{\M})$ and $X_{\oneD} = \bar{X} \equiv \frac{1}{2} (X_{\E} + X_{\M})$, let us reexpress the RHS in terms of the terminator averages $\bar{H}$ and $\bar{X}$, along with the deviations $\Delta H \equiv \frac{1}{2} (H_{\E} - H_{\M})$ and $\Delta X \equiv \frac{1}{2} (X_{\E} - X_{\M})$, giving %(such that $H_{\M} = \bar{H} - \Delta H$ and $H_{\E} = \bar{H} + \Delta H$) 
\begin{align} \label{eq:A7}
\mathrm{RHS} = \frac{\RptwoD^{2} - \RponeD^{2}}{\RponeD} \, + \frac{\RptwoD}{\RponeD} & \left\{ (\bar{H} - \Delta H) \left[ \gamma + \ln{\left( \frac{P_0}{\mu g} \sqrt{2 \pi \RptwoD} \, \sigma_{\lambda} \right)} + \ln{(\bar{H} - \Delta H)^{-\frac{1}{2}}} + \ln{(\bar{X} - \Delta X)} \right] + \right. \nonumber \\
& \hspace{8pt} \left. (\bar{H} + \Delta H) \left[ \gamma + \ln{\left( \frac{P_0}{\mu g} \sqrt{2 \pi \RptwoD} \, \sigma_{\lambda} \right)} + \ln{(\bar{H} + \Delta H)^{-\frac{1}{2}}} + \ln{(\bar{X} + \Delta X)} \right] \right\}
\end{align} 
Extracting a factor of $\bar{H}^{-\frac{1}{2}}$ and $\bar{X}$ from the second and third logarithms in each pair to subsume into the first logarithms, we can use Equation~\ref{eq:A6} to simplify the RHS
\begin{align} \label{eq:A8}
\mathrm{RHS} = \frac{\RptwoD^{2} - \RponeD^{2}}{\RponeD} \, + \frac{\RptwoD}{\RponeD} & \left\{ (\bar{H} - \Delta H) \left[ \gamma + \ln{\bar{\tau}_{0, \lambda}} + \ln{\left(1 - \frac{\Delta H}{\bar{H}}\right)^{-\frac{1}{2}}} + \ln{\left(1 - \frac{\Delta X}{\bar{X}}\right)} \right] + \right. \nonumber \\
& \hspace{8pt} \left. (\bar{H} + \Delta H) \left[ \gamma + \ln{\bar{\tau}_{0, \lambda}} + \ln{\left(1 + \frac{\Delta H}{\bar{H}}\right)^{-\frac{1}{2}}} + \ln{\left(1 + \frac{\Delta X}{\bar{X}}\right)} \right] \right\}
\end{align} 
Carrying out the addition between the two square brackets causes pairwise cancellation of some $\Delta H$ terms
\begin{align} \label{eq:A9}
\mathrm{RHS} = \frac{\RptwoD^{2} - \RponeD^{2}}{\RponeD} \, + \frac{\RptwoD}{\RponeD} & \left\{ 2 \bar{H} (\gamma + \ln{\bar{\tau}_{0, \lambda}}) + (\bar{H} - \Delta H) \left[ \ln{\left(1 - \frac{\Delta H}{\bar{H}}\right)^{-\frac{1}{2}}} + \ln{\left(1 - \frac{\Delta X}{\bar{X}}\right)} \right] + \right. \nonumber \\
& \hspace{85pt} \left. (\bar{H} + \Delta H) \left[ \ln{\left(1 + \frac{\Delta H}{\bar{H}}\right)^{-\frac{1}{2}}} + \ln{\left(1 + \frac{\Delta X}{\bar{X}}\right)} \right] \right\}
\end{align} 
Pulling out a factor of $2 \bar{H}$ from the braces allows the RHS to be succinctly written as
\begin{equation} \label{eq:A10}
\mathrm{RHS} = \frac{\RptwoD^{2} - \RponeD^{2}}{\RponeD} \, + \frac{\RptwoD}{\RponeD} 2 \bar{H} \left\{ \gamma + \ln{\bar{\tau}_{0, \lambda}} + \tilde{f} \left(\frac{\Delta H}{\bar{H}}\right) + \tilde{g} \left(\frac{\Delta H}{\bar{H}}, \, \frac{\Delta X}{\bar{X}} \right) \right\}
\end{equation} 
where $\tilde{f}$ and $\tilde{g}$ are dimensionless functions of the temperature and compositional differences between the terminators
\begin{align}
\tilde{f} \left(\frac{\Delta H}{\bar{H}}\right) = - & \frac{1}{4} \left[ \left(1-\frac{\Delta H}{\bar{H}}\right) \ln{\left(1-\frac{\Delta H}{\bar{H}}\right)} + \left(1 +\frac{\Delta H}{\bar{H}}\right) \ln{\left(1+\frac{\Delta H}{\bar{H}}\right)} \right] \label{A11} \\
\tilde{g} \left(\frac{\Delta H}{\bar{H}}, \, \frac{\Delta X}{\bar{X}} \right) = & \frac{1}{2} \left[ \left(1-\frac{\Delta H}{\bar{H}}\right) \ln{\left( 1-\frac{\Delta X}{\bar{X}}\right)} + \left( 1 +\frac{\Delta H}{\bar{H}}\right) \ln{\left(1+\frac{\Delta X}{\bar{X}}\right)} \right] \label{eq:A12}
\end{align}

We show in Appendix~\ref{appendix_B} that it is properties of $\tilde{f}$ and $\tilde{g}$ which are responsible for non-uniform terminator biases. Note that while temperature differences are expected to be small, such that $\Delta T / \bar{T} = \Delta H / \bar{H} \ll 1$, mixing ratios can differ by orders of magnitude between terminators. It is therefore more informative to consider logarithmic mixing ratios. Defining $\overline{\ln{X}} \equiv \frac{1}{2} (\ln{X_{\E}} + \ln{X_{\M}})$ and $\Delta \ln{X} \equiv \frac{1}{2} (\ln{X_{\E}} - \ln{X_{\M}})$, one can show that
\begin{align}
\Delta X = \frac{1}{2} \left(e^{\overline{\ln{X}} + \Delta \ln{X}} - e^{\overline{\ln{X}} - \Delta \ln{X}} \right) & = e^{\overline{\ln{X}}} \sinh{(\Delta \ln{X})} \label{eq:A13} \\
\bar{X} = \frac{1}{2} \left(e^{\overline{\ln{X}} + \Delta \ln{X}} + e^{\overline{\ln{X}} - \Delta \ln{X}} \right) & = e^{\overline{\ln{X}}} \cosh{(\Delta \ln{X})} \label{eq:A14}
\end{align}
Substituting $\Delta X / \bar{X} = \tanh{(\Delta \ln{X})}$ into Equation~\ref{eq:A12}, we have
\begin{equation} \label{eq:A15}
\tilde{g} = \frac{1}{2} \left[ \left(1-\frac{\Delta H}{\bar{H}}\right) \ln{\left(1 - \tanh{(\Delta \ln{X})} \right)} + \left( 1 +\frac{\Delta H}{\bar{H}}\right) \ln{\left(1 + \tanh{(\Delta \ln{X})} \right)} \right]   
\end{equation}
or
\begin{equation} \label{eq:A16}
\tilde{g} = \frac{1}{2} \left[ \ln{\left(1 - \tanh^2{(\Delta \ln{X})} \right)} + \left(\frac{\Delta H}{\bar{H}}\right) \ln{\left(\frac{1 + \tanh{(\Delta \ln{X})}}{1 - \tanh{(\Delta \ln{X})}} \right)} \right]
\end{equation}
Using the identities $1 - \tanh^2{x} = (\cosh{x})^{-2}$ and $\frac{1 + \tanh{x}}{1 - \tanh{x}} = e^{2x}$, we can finally write $\tilde{g}$ as
\begin{equation} \label{eq:A17}
\tilde{g} \left(\frac{\Delta H}{\bar{H}}, \, \Delta \ln{X} \right) = \left(\frac{\Delta H}{\bar{H}}\right) \Delta\ln{X} - \ln{\left[\cosh(\Delta\ln{X})\right]}
\end{equation}
Note that $\tilde{g}$ depends on the difference in logarithmic mixing ratios between the terminators, but not on their average.

Returning now to the transit depth equivalence condition, we can equate Equations~\ref{eq:A5} and \ref{eq:A10} to write
\begin{equation} \label{eq:A18}
2 H_{\oneD} \left[ \gamma + \ln{\bar{\tau}_{0, \lambda}} + \ln{\sqrt{\frac{\bar{H}}{H_{\oneD}}}} + \ln{\left(\frac{X_{\oneD}}{\bar{X}}\right)} + \ln{\sqrt{\frac{\RponeD}{\RptwoD}}} \right] =  \frac{\RptwoD^{2} - \RponeD^{2}}{\RponeD} \, + \frac{\RptwoD}{\RponeD} 2 \bar{H} \left[ \gamma + \ln{\bar{\tau}_{0, \lambda}} + \tilde{f} + \tilde{g} \right]
\end{equation} 
Introducing three new variables
\begin{equation} \label{eq:A19}
a = \gamma + \ln{\bar{\tau}_{0, \lambda}} + \ln{\sqrt{\bar{H}}} + \ln{\left(\frac{X_{\oneD}}{\bar{X}}\right)} + \ln{\sqrt{\frac{\RponeD}{\RptwoD}}}
\end{equation} 
\begin{equation} \label{eq:A20}
b = \frac{\RptwoD}{\RponeD} \left[\gamma + \ln{\bar{\tau}_{0, \lambda}} + \tilde{f} + \tilde{g} \right]
\end{equation} 
\begin{equation} \label{eq:A21}
c = \frac{\RptwoD^{2} - \RponeD^{2}}{\bar{H} \RponeD}
\end{equation} 
Equation~\ref{eq:A18} can be simply written as
\begin{equation} \label{eq:A22}
2 H_{\oneD} \left(a - \frac{1}{2} \ln{H_{\oneD}}\right) = \bar{H} (c + 2b)
\end{equation}
With a change of variables to $x = \ln{H_{\oneD}} - 2 a$, this becomes
\begin{equation} \label{eq:A23}
-x e^{x + 2a} = \bar{H} (c + 2b)
\end{equation}
hence
\begin{equation} \label{eq:A24}
x e^{x}  = - e^{-2a} \bar{H} (c + 2b)
\end{equation}
The solution to the equation $x e^x = y$ is given by $x = W(y)$, where $W$ is the \emph{Lambert W function} \citep{Corless1996}. We can therefore write 
\begin{equation} \label{eq:A25}
x = W \left(- e^{-2a} \bar{H} (c + 2b)\right)
\end{equation}
Hence via the definition of $x$,
\begin{equation} \label{eq:A26}
H_{\oneD} = e^{2a} e^{W \left(- e^{-2a} \bar{H} (c + 2b)\right)}
\end{equation}
The Lambert W function satisfies the property $W(y) e^{W(y)} = y$, or equivalently $e^{W(y)} = \frac{y}{W(y)}$, hence
\begin{equation} \label{eq:A27}
H_{\oneD} = \frac{e^{2a} \left(- e^{-2a} \bar{H} (c + 2b)\right)}{W \left(- e^{-2a} \bar{H} (c + 2b)\right)} = \bar{H} \left[ \frac{- (c + 2b)}{W \left(- (c + 2b) e^{-2a} \bar{H}\right)} \right]
\end{equation}
One ambiguity stems from $W(y)$ having two real branches: $W_0 (y) \geq -1$ (principal) and $W_{-1} (y) \leq -1$ (lower). Since we require $H_{\oneD} > 0$, the denominator must be negative\footnote{Taking the limit of negligible terminator asymmetries, $c \rightarrow 0$ and $b \rightarrow [\gamma + \ln{\bar{\tau}_{0, \lambda}}] > 0$, hence the numerator is negative.} for all arguments and hence we need the lower branch.

Substituting back the definitions of $a$, $b$, and $c$ (Equations~\ref{eq:A19}, \ref{eq:A20}, and \ref{eq:A21}) into Equation~\ref{eq:A27} yields
\begin{equation} \label{eq:A28}
H_{\oneD} = \bar{H} \left[ \frac{ - \left( \frac{\RptwoD^{2} - \RponeD^{2}}{\bar{H} \RponeD} + 2 \frac{\RptwoD}{\RponeD} \left[\gamma + \ln{\bar{\tau}_{0, \lambda}} + \tilde{f} + \tilde{g} \right] \right)}{W_{-1} \left(- \left( \frac{\RptwoD^{2} - \RponeD^{2}}{\bar{H} \RponeD} + 2 \frac{\RptwoD}{\RponeD} \left[\gamma + \ln{\bar{\tau}_{0, \lambda}} + \tilde{f} + \tilde{g} \right] \right) e^{-2 \left[\gamma + \ln{\bar{\tau}_{0, \lambda}} + \ln{\sqrt{\bar{H}}} + \ln{\left(X_{\oneD}/\bar{X}\right)} + \ln{\sqrt{\RponeD/\RptwoD}}\right]} \bar{H} \right)} \right]
\end{equation}
which simplifies to
\begin{equation} \label{eq:A29}
H_{\oneD} = \bar{H} \left[ \frac{ - \left( \frac{\RptwoD^{2} - \RponeD^{2}}{\bar{H} \RponeD} + 2 \frac{\RptwoD}{\RponeD} \left[\gamma + \ln{\bar{\tau}_{0, \lambda}} + \tilde{f} + \tilde{g} \right] \right)}{W_{-1} \left(- \left( \frac{\RptwoD^{2} - \RponeD^{2}}{\bar{H} \RponeD} + 2 \frac{\RptwoD}{\RponeD} \left[\gamma + \ln{\bar{\tau}_{0, \lambda}} + \tilde{f} + \tilde{g} \right] \right) \left(\frac{\bar{X}}{X_{\oneD}}\right)^2 \frac{\RptwoD}{\RponeD} e^{-2 \left[\gamma + \ln{\bar{\tau}_{0, \lambda}}\right]} \right)} \right]
\end{equation}
Finally, switching from $H$ to $T$, we arrive at the general expression for the equivalent 1D temperature
\begin{equation} \label{eq:A30}
T_{\oneD} = \bar{T} \left[ \frac{ \frac{\RptwoD}{\RponeD} \Psi_{\lambda} - \frac{\RptwoD^{2} - \RponeD^{2}}{\bar{H} \RponeD}}{W_{-1} \left(\left[ \frac{\RptwoD}{\RponeD} \Psi_{\lambda} - \frac{\RptwoD^{2} - \RponeD^{2}}{\bar{H} \RponeD} \right] \left(\frac{\bar{X}}{X_{\oneD}}\right)^2 \frac{\RptwoD}{\RponeD} e^{-2 \left[\gamma + \ln{\bar{\tau}_{0, \lambda}}\right]} \right)} \right]
\end{equation}
where we have defined
\begin{equation}  \label{eq:A31}
\Psi_{\lambda} = -2 \left[ \gamma + \ln{\bar{\tau}_{0, \lambda}} + \tilde{f}\left(\frac{\Delta T} {\bar{T}}\right) + \tilde{g}\left(\frac{\Delta T}{\bar{T}}, \Delta\ln{X}\right) \right]
\end{equation}
In the special case where the deep radius is correctly retrieved (i.e. $\RponeD = \RptwoD$), we derive Equation~\ref{eq:analytic_cooling_bias}.
\begin{equation} \label{eq:A32}
T_{\oneD} = \bar{T} \left[ \frac{\Psi_{\lambda}}{W_{-1} \left( \Psi_{\lambda} \left(\bar{X}/X_{\oneD}\right)^2 e^{-2 (\gamma + \ln{\bar{\tau}_{0, \lambda}})}  \right) } \right] 
\end{equation} 

\vspace{0.5cm}

\section{Properties of the analytic solution} \label{appendix_B}

\restartappendixnumbering

Here we demonstrate several mathematical properties of Equation~\ref{eq:analytic_cooling_bias} which give rise to retrieved temperature biases. We focus on the case where $\bar{X} = X_{\oneD}$ and $\RponeD = \RptwoD$, showing that even if mixing ratios and radii are correctly retrieved, the terminator temperature is not. 

\subsection{Recovering the uniform limit}

In the limit where $\frac{\Delta T}{\bar{T}}, \, \Delta\ln{X} \rightarrow 0$, we have a uniform, 1D, atmosphere. We therefore expect $T_{\oneD} \rightarrow \bar{T}$. In this limit, Equations~\ref{eq:f} and \ref{eq:g} give $\tilde{f}, \, \tilde{g} \rightarrow 0$, hence Equation~\ref{eq:analytic_cooling_bias} becomes
\begin{equation} \label{eq:B1}
T_{\oneD} = \bar{T} \left[ \frac{\Psi_{\lambda}^{'}}{W_{-1} \left( \Psi_{\lambda}^{'} e^{\Psi_{\lambda}^{'}}  \right) } \right] = \bar{T}
\end{equation} 
where $\Psi_{\lambda}^{'} = -2 \left[ \gamma + \ln{\bar{\tau}_{0, \lambda}} \right]$ and the last equality uses $W (x e^x) = x$, following from the definition of the W function.

\subsection{Pure temperature difference biases}

When $\Delta\ln{X} = 0$, we have $\tilde{g} = 0$ hence Equation~\ref{eq:analytic_cooling_bias} becomes
\begin{equation} \label{eq:B2}
T_{\oneD} = \bar{T} \left[ \frac{-2 \left[ \gamma + \ln{\bar{\tau}_{0, \lambda}} + \tilde{f}\left(\frac{\Delta T} {\bar{T}}\right) \right]}{W_{-1} \left( -2 \left[ \gamma + \ln{\bar{\tau}_{0, \lambda}} + \tilde{f}\left(\frac{\Delta T} {\bar{T}}\right) \right] e^{-2 (\gamma + \ln{\bar{\tau}_{0, \lambda}})}  \right) } \right] 
\end{equation} 
Considering a hot Jupiter with $\bar{T} = 1400\,$K and a typical morning-evening temperature difference $T_{\E} - T_{\M} \approx 200\,$K \citep{Kataria2016}, we have $\Delta T / \bar{T} \approx 0.07$. As $\Delta T / \bar{T} \ll 1$, we can Taylor expand Equation~\ref{eq:f} to yield
\begin{equation} \label{eq:B3}
\tilde{f} = -\frac{1}{4} \left[ \left(\frac{\Delta T}{\bar{T}}\right)^2 + \mathcal{O} \left(\frac{\Delta T}{\bar{T}}\right)^4 + \dots \, \right]
\end{equation} 
where the odd terms cancel due to symmetry. $\tilde{f}$ is then essentially a minor quadratic perturbing term in Equation~\ref{eq:B2}. As $\bar{\tau}_{0, \lambda}$ is defined in the deep atmosphere at $P = P_0$ (Equation~\ref{eq:tau_0}), we have $\gamma + \ln{\bar{\tau}_{0, \lambda}} \gg \tilde{f}$ and hence Equation~\ref{eq:B2} tends towards Equation~\ref{eq:B1} and $T_{\oneD} \approx \bar{T}$. Numerical exploration of Equation~\ref{eq:B2} yields cooling biases of $\lesssim 5\,$K, even for ultra-hot Jupiters. We conclude that pure temperature differences have little effect on retrieved temperatures.

\subsection{Compositional difference biases}

To consider the effect of compositional differences, let us write Equation~\ref{eq:analytic_cooling_bias} with $\tilde{f} \approx 0$ (as shown in the last section) 
\begin{equation} \label{eq:B4}
T_{\oneD} \approx \bar{T} \left[ \frac{-2 \left[ \gamma + \ln{\bar{\tau}_{0, \lambda}} + \tilde{g}\left(\frac{\Delta T}{\bar{T}}, \Delta\ln{X}\right) \right]}{W_{-1} \left( -2 \left[ \gamma + \ln{\bar{\tau}_{0, \lambda}} + \tilde{g}\left(\frac{\Delta T}{\bar{T}}, \Delta\ln{X}\right) \right] e^{-2 (\gamma + \ln{\bar{\tau}_{0, \lambda}})}  \right) } \right] 
\end{equation} 
Writing the denominator as $W_{-1} (y)$, the exponential factor implies $|y| \ll 0.1$. We can then employ an asymptotic expansion, valid for $-0.1 \leq y \leq 0$, to write \citep{Vazquez-Leal2019}
\begin{equation} \label{eq:B5}
W_{-1} (y) \approx \ln{(-y)} - \ln{(-\ln{(-y)})} + \frac{\ln{(-\ln{(-y)})}}{\ln{(-y)}}
\end{equation}
Due to the logarithmic dependence of $W_{-1} (y)$ on $y$, whilst $y$ itself varies roughly linearly in $\Delta\ln{X}$, to a zeroth approximation we can take $W_{-1} (y) \approx - \mathrm{const}$. This allows a simpler functional form to be written
\begin{equation} \label{eq:B6}
T_{\oneD} \appropto \bar{T} \left[ \gamma + \ln{\bar{\tau}_{0, \lambda}} + \tilde{g}\left(\frac{\Delta T}{\bar{T}}, \Delta\ln{X}\right) \right]
\end{equation} 
Taking the limit $e^{\Delta \ln{X}} \gg e^{-\Delta \ln{X}}$, for which $\tilde{g} \rightarrow \ln{2} - \left(1 - \frac{\Delta T}{\bar{T}}\right) \Delta \ln{X}$, the asymptotic behaviour of compositional differences is
\begin{equation} \label{eq:B7}
T_{\oneD} \appropto \bar{T} \left[ -\left(1 - \frac{\Delta T}{\bar{T}}\right) \Delta \ln{X} + \left( \gamma + \ln{\bar{\tau}_{0, \lambda}} + \ln{2} \right) \right]
\end{equation} 
which explains the roughly linear decrease of $T_{\oneD}$ with $\Delta \log_{10} (X)$ shown in Figure~\ref{fig:analytic_solution}. The factor of $\left(1 - \frac{\Delta T}{\bar{T}}\right)$ modulating the gradient also explains why the coldest values of $T_{\oneD}$ occur for pure compositional differences.

Finally, we note that the condition for cooling biases to occur from Equation~\ref{eq:B4} can be essentially reduced to $\tilde{g} < 0$. However, in the presence of both temperature and compositional differences, Equation~\ref{eq:g} has a regime where $\tilde{g} > 0$ (the `wrinkle' in Figure~\ref{fig:analytic_solution}). We can therefore define a `critical' mixing ratio difference, $\Delta \ln{X_{\rm{crit}}}$, where cooling biases begin according to $\tilde{g} = 0$, or
\begin{equation} \label{eq:B8}
\left(\frac{\Delta T}{\bar{T}}\right) \Delta \ln{X_{\rm{crit}}} = \ln{\left[\cosh(\Delta\ln{X_{\rm{crit}}})\right]}
\end{equation}
Numerically solving this equation results in $\Delta \ln{X_{\rm{crit}}} \appropto \frac{\Delta T}{\bar{T}}$ as long as $\frac{\Delta T}{\bar{T}} \lesssim 0.7$ (with $\Delta \ln{X_{\rm{crit}}} \rightarrow \infty$ as $\frac{\Delta T}{\bar{T}} \rightarrow 1$). To obtain a quantitative handle on the scale of $\Delta \ln{X_{\rm{crit}}}$, consider a conservative `extreme' temperature difference of $T_{\E} - T_{\M} = 1000\,$K on an ultra-hot Jupiter with $\bar{T} = 2500\,$K. This case has $\frac{\Delta T}{\bar{T}} = 500/2500 = 0.2$ and a numerical solution of $\Delta \ln{X_{\rm{crit}}} \approx 0.41$ ($\Delta \log_{10}{X_{\rm{crit}}} \approx 0.18$), or equivalently a difference of $\approx 50\,$\%. Therefore even minor differences in mixing ratios between the terminators enter the cooling bias regime. As a rule of thumb, compositional differences satisfying $\Delta \log_{10}{X} > 0.3$ (a factor of 2 difference) result in $T_{\oneD}$ biased to hundreds of K colder than $\bar{T}$.

\bibliography{coldinhere}{}
\bibliographystyle{aasjournal}

\end{document}